\documentclass[aps,prl,twocolumn,superscriptaddress,showpacs,nofootonbib]{revtex4}
\usepackage{amssymb}
\usepackage{amsmath}
\usepackage{amsthm}
\usepackage{amsfonts}
\usepackage{color}
\usepackage{amscd}
\usepackage{graphicx}
\allowdisplaybreaks

\newtheorem{prop}{Proposition}

\newtheorem{proto}{Protocol}
\newcommand{\ket}[1]{|#1\rangle}
\newcommand{\bra}[1]{\langle #1 |}

\newcommand{\nn}{\nonumber}

\newcommand{\rg}{\mathop{\rm r }\nolimits}

\begin{document}

\title{Sequential Quantum Cloning}

\author{Y. Delgado}
 \affiliation{Secci\'{o}n
F\'{\i}sica, Departamento de Ciencias, Pontificia Universidad
Cat\'{o}lica del Per\'{u}, Apartado Postal 1761, Lima, Peru}

\author{L. Lamata}
 \affiliation{Instituto de
Matem\'{a}ticas y F\'{\i}sica Fundamental, CSIC, Serrano 113-bis,
28006 Madrid, Spain}

\author{J. Le\'{o}n}
 \affiliation{Instituto de
Matem\'{a}ticas y F\'{\i}sica Fundamental, CSIC, Serrano 113-bis,
28006 Madrid, Spain}

\author{D. Salgado}
 \affiliation{Dpto.\ F\'{\i}sica
Te\'{o}rica, Universidad Aut\'{o}noma de Madrid, 28049 Cantoblanco,
Madrid, Spain}

\author{E. Solano}
\affiliation{Secci\'{o}n
F\'{\i}sica, Departamento de Ciencias, Pontificia Universidad
Cat\'{o}lica del Per\'{u}, Apartado Postal 1761, Lima, Peru}
\affiliation{Max-Planck-Institut f\"ur Quantenoptik,
Hans-Kopfermann-Strasse 1, 85748 Garching, Germany}
\affiliation{Physics Department, ASC, and CeNS,
Ludwig-Maximilians-Universit\"at, Theresienstrasse 37, 80333 Munich,
Germany}

\date{\today}

\begin{abstract}
Not all unitary operations upon a set of qubits can be implemented
by sequential interactions between each qubit and an ancillary
system. We analyze the specific case of sequential quantum
cloning, $1 \to M$, and prove that the minimal dimension $D$ of
the ancilla grows \emph{linearly} with the number of clones $M$.
In particular, we obtain $D=2M$ for symmetric universal quantum
cloning and $D=M+1$ for symmetric phase-covariant cloning.
Furthermore, we provide a recipe for the required ancilla-qubit
interactions in each step of the sequential procedure for both
cases.
\end{abstract}


\maketitle

Multipartite entangled states stand up as the most versatile and
powerful tool to perform information-processing protocols in
Quantum Information Science \cite{BenDiV00a}. They arise as an
invaluable resource in tasks such as quantum computation
\cite{DeuEke98a,RauBri01a}, quantum state teleportation
\cite{BouEkeZei00a}, quantum communication \cite{HorHorHor01a} and
dense coding \cite{BenWie92a}. As a result, the controllable
generation of these states becomes a crucial issue in the quest
for quantum-informational proposals. However, the generation of
multipartite entangled states through single global unitary
operations is, in general, an extremely difficult experimental
task. In this sense, the sequential generation studied by
Sch\"{o}n \textit{et al.}\ \cite{SchSolVerCirWol05a}, where at
each step one qubit is allowed to interact with an ancilla,
appears as the most promising avenue. The essence of this
sequential scheme is the successive interaction of each qubit
initialized in the standard state $\ket{0}$ with an ancilla of a
suitable dimension $D$ to generate the desired multiqubit state.
In the last step, the qubit-ancilla interaction is chosen so as to
decouple the final multiqubit entangled state from the auxiliary
$D$-dimensional system, yielding~\cite{SchSolVerCirWol05a}
\begin{equation}\label{GenMPS}
\ket{\Psi}=\sum_{i_{1}\cdots
i_{n}=0,1}\bra{\varphi_{F}}V_{[n]}^{i_{n}} \cdots
V_{[1]}^{i_{1}}\ket{\varphi_{I}}\ket{i_{1}\cdots i_{n}} .
\end{equation}
Here, the $V_{[k]}^{i_{k}}$ are $D-$dimensional matrices arising
from the isometries (unitaries) $V_{[k]}:\mathfrak{h}_{A}\otimes (
\ket{0} ) \to\mathfrak{h}_{A}\otimes\mathfrak{h}_{B_{k}}$, with
$\mathfrak{h}_{A}=\mathbb{C}^{D}$ and
$\mathfrak{h}_{B_{k}}=\mathbb{C}^{2}$ being the Hilbert spaces for
the ancilla and the $k$th qubit, respectively, and where
$\ket{\varphi_{I}}$ and $\ket{\varphi_{F}}$ denote the initial and
final states of the ancilla, respectively. The state
\eqref{GenMPS} is, indeed, a Matrix Product State (MPS) (cf.\
e.g.\ \cite{Eck05a} and references therein), already present in
spin chains \cite{AffKenLieTas87a}, classical simulations of
quantum entangled systems \cite{Vid03a} and density-matrix
renormalization group techniques \cite{VerPorCir04a}. Moreover, it
was proven that any multiqubit MPS can be sequentially generated
using the recipe of Ref.~\cite{SchSolVerCirWol05a}. Notice that in
this formalism, the mutual qubit-ancilla interaction in each step
$k$ completely determines the matrices $V_{[k]}^{i_{k}}$,
$i_{k}=0,1$, whereas we enjoy some freedom to build such an
interaction from a known $V_{[k]}^{i_{k}}$. This freedom stems
from the fact that in the proposed scheme only the initial state
$\ket{0}$ for each qubit is relevant.

In this letter, we consider the possibility of implementing
quantum cloning based on a sequential protocol with the help of an
ancillary system. This problem is certainly far from being an
application of Ref.~\cite{SchSolVerCirWol05a}, given that the
initial and final states are unknown. In this sense, any proposed
strategy will be closer to the open problem of which global
unitary operations (certainly not all of them) can be implemented
through a sequential procedure. Despite the fundamental no-cloning
theorem \cite{WooZur82a}, stating the impossibility to exactly
clone an unknown quantum state, there exists several cloning
techniques with a given optimal fidelity \cite{ScaIblGisAci05a}.
These procedures differ either from the initial set of states to
be cloned or from symmetry considerations. In general, an
optimality condition of the cloning procedure is obtained via the
maximization of the fidelity between the original qubit and each
final clone state. We will show how to perform sequentially both
the universal symmetric \cite{BuzHil96a,GisMas97a} and the
economical phase-covariant symmetric quantum cloning
\cite{DArMac03a,BusDArMac05a} from one qubit to $M$ clones. In the
first case, a global unitary evolution transforms \emph{any input
state} $\ket{\psi}$ in a set of $M$ clones whose individual
reduced states $\rho_{out}$ carry maximal fidelity with respect to
$\ket{\psi}$: $F_{1,M}=\frac{2M+1}{3M}$. This cloning procedure is
fully described by the evolution
\begin{eqnarray}
&& \!\!\!\!\!\!\!
\ket{\psi}\otimes\ket{B}\to\ket{GM_{M}(\psi)}\equiv \label{GisinMassar}
 \nonumber \\ && \equiv \!\! \sum_{j=0}^{M-1} \!\! \alpha_{j}\ket{(M-j)\psi,j\psi^{\perp}}_{S}
\! \otimes \! \ket{(M-j-1)\psi^{*},j\psi^{*\perp}}_{S} , \nonumber \\
\end{eqnarray}
\noindent where $\ket{GM_{M}(\psi)}$ stands for the state produced by the
Gisin-Massar cloning procedure~\cite{GisMas97a}, that results in $M$ optimal clones of
$\ket{\psi}$ from the initial blank state $\ket{B}$,
$\alpha_{j}=\sqrt{\frac{2(M-j)}{M(M+1)}}$, and
$\ket{(M-j)\psi,j\psi^\perp}_{S}$ denotes the normalized completely
symmetric state with $(M-j)$ qubits in state $\psi$ and $j$ qubits
in state $\psi^{\perp}$. Notice the presence of $M-1$ additional
so-called anticlones. They are
necessary in order to perform this cloning procedure with the
optimal fidelity. The anticlone state $\psi^{*}$ refers to the fact
that they transform under rotations as the complex conjugate
representation. For concreteness sake we have chosen
$\ket{\psi^{*}}=\cos\theta/2\ket{1}+e^{-i\phi}\sin\theta/2\ket{0}$
in coincidence with the seminal paper by Bu\v{z}ek and Hillery
\cite{BuzHil96a}, whereas
$\ket{\psi}=\cos\theta/2\ket{0}+e^{i\phi}\sin\theta/2\ket{1}$. In
the second case, motivated by quantum cryptoanalysis, the goal is to
clone only those states belonging to the equatorial plane of the
Bloch sphere, i.e.\ those such that $\theta=\pi/2$. Furthermore, we
have only focused upon the cases where no anticlones are needed
(hence the term economical). Under this assumption, imposing the
purity of the joint state, the number of clones $M$ must be odd
\cite{DArMac03a}. The cloning evolution is now given by
\begin{equation}\ket{\psi}\otimes\ket{B}
\to\frac{1}{\sqrt{2}}\left[\ket{(k+1)0,k1}_{S}+e^{i\phi}
\ket{k0,(k+1)1}_{S}\right],\label{DArianoMacchia}
\end{equation}
\noindent where $k=(M-1)/2$ and where we have followed the same
convention as above.

The basic idea is to express the final states \eqref{GisinMassar} and
\eqref{DArianoMacchia} in its MPS form, as given in Ref.~\cite{Vid03a},
by performing $n-1$ sequential
Schmidt decompositions $$\ket{\Phi}=\sum_{\alpha_{1}\dots
\alpha_{n-1}} \ket{\varphi^{[1]}_{\alpha_1}}\lambda[1]_{\alpha_{1}}
\ket{\varphi^{[2]}_{\alpha_1 \alpha_2}} \cdots \lambda[n-1]_{\alpha_{n-1}}
\ket{\varphi^{[n]}_{\alpha_{n-1}}},$$ and then writing the unnormalized
Schmidt states
in the computational basis for the corresponding qubit
$\ket{\varphi^{[l]}_{\alpha_{l-1}\alpha_l}}=\sum_l
\Gamma[l]^{i_{l}}_{\alpha_{l-1}\alpha_{l}} \ket{i_l}$. Then,
$\ket{\Phi}=\sum_{i_{1}\dots
i_{N}}c_{i_{1}\dots i_{N}}\ket{i_{1}\dots i_{N}}$, with
\begin{equation}\label{Vidal}c_{i_{1}\dots i_{N}}=\sum_{\alpha_{1}
\dots\alpha_{n-1}}
\Gamma[1]^{i_{1}}_{\alpha_{1}}\lambda[1]_{\alpha_{1}}
\Gamma[2]^{i_{2}}_{\alpha_{1}\alpha_{2}}
\lambda[2]_{\alpha_{2}}\dots\Gamma[n]_{\alpha_{n-1}}^{i_{n}}.
\end{equation}
We identify the matrices $V^{i_{k}}_{[k]}$ by matching indices in
expressions \eqref{GenMPS} and \eqref{Vidal}. The indices
$\alpha_{j}$ run from $1$ to $\chi$, where
$\chi=\max_{\mathcal{P}}\{\chi_{\mathcal{P}}\}$,
$\chi_{\mathcal{P}}$ denoting the rank of the reduced density
matrix $\rho_{\mathcal{P}}$ for the bipartite partition
$\mathcal{P}$ of the composite system \cite{Vid03a}.

In order to employ the sequential
ancilla-qubit device as a quantum cloning machine we will firstly
elucidate the minimal dimension required for the ancilla. To clone an
arbitrary input qubit state
$\ket{\psi}=\alpha\ket{0}+\beta\ket{1}$, we exploit linearity and
determine the minimal dimension $D_{0,1}$ of the ancillas to
perform the cloning for the states $\ket{0}$ and  $\ket{1}$ and then
 combine both results in a single
ancilla of minimal dimension $D$ to be determined.
Let us focus upon the symmetric universal cloning of
$\ket{0}$. To determine the minimal dimension $D_{0}$ of the
ancilla we need to compute $\chi$, which can be done without the
exact MPS expression for the state.

Let us denote by $\mathcal{P}=A|B$ the partition into two subsystems,
one with the first $A$ qubits, the other with the following $B$
qubits, and $C_{A|B}$ the corresponding coefficient matrix. For
definiteness, $C_{A|B}(\psi)=[c_{i_1\dots i_A, \,i_{A+1}\dots i_{A+B}}]$,
where $i_1\dots i_A$ is treated as the row index, whereas
$i_{A+1}\dots i_{A+B}$ is treated as the column index, and
 $c_{i_1\dots i_A,i_{A+1}\dots i_{A+B}}$ denote the coefficients of
 state $\ket{\psi}$. Now, the Gisin-Massar   state  cloned from $\ket{0}$ can be written as
\begin{eqnarray}
&&\!\!\!\!\!\!\! \ket{GM_{M}(0)}=\label{juan1} \nonumber \\
&&\mathcal{S}_{M}\otimes \mathcal{S}_{M-1}
\!\! \sum_{j=0}^{M-1}
\alpha_{j}\ket{(M-j)0,j1}\otimes\ket{(M-j-1)1,j0}, \nonumber \\
\end{eqnarray}
where $\mathcal{S}_A$ is the normalized symmetrizing
operator for $A$ qubits, so that
$\mathcal{S}_{M}\otimes \mathcal{S}_{M-1}$ is an invertible
local operator for the partition $M|M-1$. Due to the orthonormalities
among the states on the rhs, their $C_{M|M-1}$ can only have $M$
different rows whereas the rest are all null, hence $r(C_{M|M-1})=M$.
As $\mathcal{S}_{M}\otimes \mathcal{S}_{M-1}$ amounts to local
changes of basis within both partitions only, they cannot change
the rank of the density matrix $\rho_{M|M-1}$, so that the
rank of the coefficient matrix of \eqref{GisinMassar} is also $M$.
We now consider the
partition $k|2M-k-1$, where $k=1,\dots M-2$. The matrices
$C_{k|2M-k-1}$ are obtained from the $C_{M|M-1}$ by adjoining
rows and columns to make them longer, but  -- as that there are only
$M$ different rows in $C_{M|M-1}$, the rest
being all null -- this reordering procedure cannot increase
the former rank. Finally,
\begin{equation}
\rg(C_{k|2M-1-k})\leq\rg(C_{M|M-1}) =  M.
\label{juan2}
\end{equation}

From the results above, it follows that $\chi=M$,
i.e.\ that the minimal dimension $D_{0}$ to clone the $\ket{0}$
state is $D_{0}=M$, namely the number of clones to produce.
Repeating the same argument for the initial state $\ket{1}$ we
also conclude that the minimal dimension of the ancilla to clone
the $\ket{1}$ state is $D_{1}=M$, as expected. Now we must combine
both results to find $D$ for an arbitrary unknown state
$\ket{\psi}=\alpha\ket{0}+\beta\ket{1}$. It is a wrong guessing to
think that it should also be $D=M$ and, consequently, a different
scheme must be given. The MPS expression of \eqref{GisinMassar}
for the original state $\ket{0}$ determines the $D$-dimensional
matrices $V_{0[k]}^{i_{k}}$, whereas the corresponding MPS
expression for the original state $\ket{1}$ determines
$V_{1[k]}^{i_{k}}$,

\begin{eqnarray}
&& \ket{GM_{M}(0)}=\sum_{i_{1}\dots
i_{n}=0,1}\bra{\varphi_{F}^{(0)}}
V_{0[n]}^{i_{n}}\dots V_{0[1]}^{i_{1}}\ket{0}_{D}\ket{i_{1}\dots i_{n}}, \nn \\
&& \ket{GM_{M}(1)}=\sum_{i_{1}\dots
i_{n}=0,1}\bra{\varphi_{F}^{(1)}}V_{1[n]}^{i_{n}}\dots
V_{1[1]}^{i_{1}}\ket{0}_{D}\ket{i_{1}\dots i_{n}}. \nn \\
\end{eqnarray}

Here, $\ket{\varphi_{F}^{(0)}}$ and $\ket{\varphi_{F}^{(1)}}$ can be
calculated explicitly and will play an important role below.

We propose now to double the dimension of the ancilla,
$\mathbb{C}^{D}\to\mathbb{C}^{2} \otimes\mathbb{C}^{D}$, in order
to implement a deterministic protocol of sequential quantum
cloning.
\begin{proto}
\begin{enumerate}
\item[i.] Encode the unknown state $\ket{\psi}$ in the initial
ancilla state $\ket{\varphi_{I}}=\ket{\psi}\otimes\ket{0}_{D}$.
\item[ii.] Allow each qubit $k$ to interact with the ancilla
according to the $2D$-dimensional isometries
$V_{[k]}^{i_{k}}=\ket{0}\bra{0}\otimes V_{0[k]}^{i_{k}} +
\ket{1}\bra{1}\otimes V_{1[k]}^{i_{k}}$. \item[iii.] Perform a
generalized Hadamard transformation upon the ancilla

\begin{eqnarray}
\ket{0}\otimes\ket{\varphi_{F}^{(0)}}\to\frac{1}{\sqrt{2}}\left[\ket{0}
\otimes\ket{\varphi_{F}^{(0)}}+\ket{1}\otimes\ket{\varphi_{F}^{(1)}}
\right], \nonumber \\
\ket{1}\otimes\ket{\varphi_{F}^{(1)}}\to\frac{1}{\sqrt{2}}
\left[\ket{0}\otimes\ket{\varphi_{F}^{(0)}}-\ket{1}\otimes
\ket{\varphi_{F}^{(1)}}\right] . \label{Hadamard}
\end{eqnarray}
Note that the choice $\mathbb{C}^{D}\to\mathbb{C}^{2}
\otimes\mathbb{C}^{D}$ (based on pedagogical reasons) could be
changed, equivalently, to $\mathbb{C}^{D} \to \mathbb{C}^{2D}$. In
this way, Eq.~(\ref{Hadamard}) would not display entangled states
but simple linear superpositions.

\item[iv.] Perform a measurement upon the ancilla in the local
basis $\{\ket{0}\otimes\ket{\varphi_{F}^{(0)}},\ket{1}\otimes
\ket{\varphi_{F}^{(1)}}\}$. \item[v.] If the result is
$\ket{0}\otimes\ket{\varphi_{F}^{(0)}}$ (which happens with
probability $1/2$), the qubits are already in the desired state;
if the result is $\ket{1}\otimes\ket{\varphi_{F}^{(1)}}$
(probability $1/2$), perform a local $\pi$-phase gate upon each
qubit, then they will end up in the desired state.
\end{enumerate}
\end{proto}

\begin{proof}
After the first two steps, the joint state of the ancilla and the
qubits is $\alpha\left(\ket{0}\otimes\ket{\varphi_{F}^{(0)}} \right)
\otimes\ket{GM_{M}(0)}+\beta\left(\ket{1}\otimes
\ket{\varphi_{F}^{(1)}}\right)\ket{GM_{M}(1)}$, where originally
$\ket{\psi}=\alpha\ket{0}+\beta\ket{1}$. After the Hadamard gate in
(iii), this state becomes
\begin{eqnarray}
&&\frac{1}{\sqrt{2}}\left(\ket{0}\otimes\ket{\varphi_{F}^{(0)}}
\right)\otimes\left[\alpha\ket{GM_{M}(0)}+\beta\ket{GM_{M}(1)}
\right] + \nn \\ && + \frac{1}{\sqrt{2}}\left(\ket{1}
\otimes\ket{\varphi_{F}^{(1)}} \right) \otimes
\left[\alpha\ket{GM_{M}(0)} - \beta\ket{GM_{M}(1)}\right].\nn
\end{eqnarray}
The remaining steps follow immediately from this expresion and
from linearity \cite{GisMas97a}.
\end{proof}

Notice that despite the measurement process in step (iv), the
desired state is obtained with probability $1$, while the fidelity
of each clone is optimal, $F_{1,M}=\frac{2M+1}{3M}$, as in
Ref.~\cite{GisMas97a}. In summary, the minimal dimension $D$ of the
ancilla for cloning $M$ qubits is $D=2\times M$, i.e., it grows
linearly with the number of clones even if their Hilbert space grows
exponentially ($2^{M}$).

\vspace*{-6mm}
\begin{widetext}
\begin{center}
\begin{table}[h!]
\begin{tabular}{c|c|c}
 & $k=0$ & $k=1$ \\\hline
 &&\\
 $\left[V_{0[1]}^{k}\right]_{ij}=$ & $\left\{\begin{array}{cc}
\delta_{ij}\mathcal{C}(2-i,i-1)& 1\leq i,j\leq 2\\
\frac{1}{\sqrt{2}}\delta_{ij}& \textrm{otherwise}
\end{array}\right.$ & $\left\{\begin{array}{cc}
\delta_{i, 3-j}\mathcal{C}(2-i,i-1)& 1\leq i,j\leq 2\\
\frac{1}{\sqrt{2}}\delta_{ij} & \textrm{otherwise}
\end{array}\right. $\\
&&\\\hline &&\\
 $\left[V_{0[n]}^{k}\right]_{ij}=$ &
$\left\{\begin{array}{cc}
\delta_{ij}\frac{\mathcal{C}(n+1-i,i-1)}{\mathcal{C}(n-i,i-1)}
& 1\leq i, j\leq n\\
\frac{1}{\sqrt{2}}\delta_{ij} & \textrm{otherwise}
\end{array}\right.$& $\left\{\begin{array}{cc}
\frac{1}{\sqrt{2}} & i=1;j=n+1\\
\delta_{i,j+1}\frac{\mathcal{C}(n-j,j)}{\mathcal{C}(n-j,j-1)} &
2\leq i\leq n+1; 1\leq j\leq n\\
\frac{1}{\sqrt{2}}\delta_{ij}&
\textrm{otherwise}\end{array}\right.$\\
&&\\\hline &&\\
$\left[V_{0[M]}^{k}\right]_{ij}=$ & $\left\{\begin{array}{cc}
\delta_{ij}\frac{\alpha_{i-1}}{\mathcal{C}(M-i,i-1)
\sqrt{\binom{M}{i-1}}}&1\leq i,j\leq M\end{array}\right.$ &
$\left\{\begin{array}{cc}
\delta_{i,j+1}\frac{\alpha_{j}}{\mathcal{C}(M-j,j-1)\sqrt{\binom{M}{j}}}
& 1\leq i,j\leq M\\
\end{array}\right.$\\
&&\\\hline &&\\
 $\left[V_{0[M+n]}^{k}\right]_{ij}=$ &
$\left\{\begin{array}{cc} \delta_{i,j-1}\sqrt{\frac{i}{M-n}} &
\left\{\begin{array}{c}1\leq
i\leq M-n\\ 2\leq j\leq M-n+1\end{array}\right.\\
0 & i=M-n+1;1\leq j\leq M\\
\frac{1}{\sqrt{2}}\delta_{ij}& \textrm{otherwise}\end{array}\right.$
&$\left\{\begin{array}{cc} \delta_{ij}\sqrt{\frac{M-n+1-i}{M-n}} &
1\leq i,j\leq M-n \\0 & i=M-n+1;1\leq j\leq M\\
\frac{1}{\sqrt{2}}\delta_{ij}&
\textrm{otherwise}\end{array}\right.$\\
\end{tabular}
\caption{Matrices for the universal symmetric cloning protocol.
\label{Table}}
\end{table}
\end{center}
\end{widetext}

It can be checked straightforwardly that if one had to clone a
$d$-dimensional system, the minimal dimension for the ancilla
would be $D=d\times M$, an obvious generalization of the preceding
argument.

For the symmetric phase-covariant cloning, the same arguments can
be reproduced. For example, the first term on the r.h.s. of
Eq.~(\ref{DArianoMacchia}) can be cast in the form of the state in
Eq.~\eqref{GisinMassar}
\begin{eqnarray}
&& \!\!\!\!\!\!\!\!\!\!\!\!\!\!\! \ket{(k+1)0,k1}_{S} = \nonumber\\
&& = \sum_{j=0}^{k}\gamma_{j}\ket{(k+1-j)0,j1}_{S}\otimes
\ket{(k-j)1,j0}_{S},
\end{eqnarray}
\noindent where $\gamma_{j}\neq 0$ for all $j$, and similarly for
the second term. Thus for symmetric phase-covariant cloning the
minimal dimension for the ancilla is $D=2\times(k+1)=2\times
\frac{M+1}{2}=M+1$. We see that the dimension of the ancilla $D$
also grows linearly with the number of clones, although it is now
lesser than above. This is a direct consequence of the reduction
in the set of possible original states to clone.\\
For the symmetric universal cloning we give in detail in Table
\ref{Table} the $2D-$dimensional matrices $V_{[k]}^{i_{k}}$ driving
us to a concrete sequential scheme, and where $\mathcal{C}(i,j)=
\sqrt{\frac{1}{\binom{i+j}{i}} \sum_{k=j}^{M-1}|\alpha_{k}|^{2}
\frac{\binom{M-k}{i}\binom{k}{j}}{\binom{M}{i+j}}}$,
$\binom{p}{q}=0$ if $q>p$ and  $1< n\leq M-1$. Furthermore, we also
have $V_{1[k]}^{i_{k}}=V_{0[k]}^{\bar{i_{k}}}$, where by $\bar{i}$
we indicate $\bar{i}= i\bigoplus 1$ (mod 2). They coincide also with
the ones for the symmetric phase-covariant cloning just by doing the
substitutions $M\rightarrow \frac{M+1}{2}$ and
$\alpha_j\rightarrow\gamma_j = \sqrt{\frac{\binom{k+1}{k+1-j}
\binom{k}{j}}{\binom{2k+1}{k+1}}}$.\\
It can be readily verified that the minimal dimension for the
ancilla is $2\times M$. When sequentially applying these matrices
to the initial state $\ket{\varphi_{I}}$ of the ancilla, one can
check, as expected, that if we were to stop at the $M$th step, the
$M$ clones would have already been produced with the desired
properties, although in a highly entangled state with the ancilla.
To arrive at a final uncoupled state, the remaining $M-1$
anticlones must be operated upon by the ancilla. Note the exponential
gain achieved with this protocol; despite the $2^{M}$-dimensional
Hilbert space of the $M$ clones, we just need a $2M$-dimensional
ancilla. This is a consequence of the Matrix-Product
decomposition of the Gisin-Massar universal cloning state.\\
The proposed schemes can be implemented in a variety of physical
setups: microwave and optical cavity QED, circuit QED, trapped ions,
and quantum dots, among others. As a paradigmatic example, the clone
could be codified in a photonic state and the ancilla in a $D$-level
atom~\cite{SchSolVerCirWol05a}, and the sequential operations
carried out by Raman lasers would produce unitaries associated with
the isometries $V^{i_k}_{[k]}$ appearing in Table I. These and other required unitary operations, as local Hadamard gates, are standard in most of the above mentioned physical setups, making our proposal suitable for future implementation.

In conclusion, we have shown how to reproduce sequentially both
the symmetric universal and symmetric phase-covariant cloning
operations. For the universal cloning we have proved that the
minimal dimension for the ancilla should be $D=2M$, where $M$
denotes the number of clones, thus showing a linear dependence.
The original state must be encoded in a $2D-$dimensional state.
For the phase-covariant case, the required dimension $D$ of the
ancilla can be reduced to $D=M+1$. In both cases, the ancilla ends
up uncoupled to the qubits. Along similar lines, this sequential
cloning protocol can be adapted to other proposals, such as
asymmetric universal quantum cloning machines or other
state-dependent protocols. This procedure can have notable
experimental interest, since it provides a systematic method to
furnish any multiqubit state using only sequential two-system
(qubit-ancilla) operations.

Y.D. thanks the support of DAI-PUCP through PAIN. L.L. acknowledges
support from FPU grant No.~AP2003-0014, L.L. and J.L. from Spanish
MEC No.~FIS2005-05304, and D.S. from MEC No.~FIS2004-01576. E.S.
thanks S. Iblisdir and J.I. Latorre for useful discussions, and
acknowledges the support of EU through RESQ and EuroSQIP, and of DFG
through SFB 631.

\end{document}